\documentclass[onecolumn,english,prd]{revtex4}
\usepackage{graphicx}
\begin{document} 

\newcommand{\lexp}{\mathop{\langle}}
\newcommand{\rexp}{\mathop{\rangle}}
\newcommand{\rexpc}{\mathop{\rangle}}
\newcommand{\beq}{\begin{equation}}
\newcommand{\eeq}{\end{equation}}
\newcommand{\beqa}{\begin{eqnarray}}
\newcommand{\eeqa}{\end{eqnarray}}

\def\k{{\hbox{\bf k}}}
\def\q{{\hbox{\bf q}}}
\def\x{{\hbox{\bf x}}}
\def\p{{\hbox{\bf p}}}
\def\dD{\delta_{\rm D}}

\def\Mpc{\, h^{-1} \, {\rm Mpc}}
\def\Gpc{\, h^{-1} \, {\rm Gpc}}
\def\Gpccube{\, h^{-3} \, {\rm Gpc}^3}
\def\kvecMpc{\, h \, {\rm Mpc}^{-1}}

\title{Memory of Initial Conditions in Gravitational Clustering}
\author{Mart\'{\i}n Crocce and  Rom\'an Scoccimarro}
\vskip 2pc
\address{Center for Cosmology and Particle Physics,  \\
Department of Physics, New York University, New York, NY 10003}

\begin{abstract}

We study the nonlinear propagator, a key ingredient in renormalized perturbation theory (RPT) that allows a well-controlled extension of perturbation theory into the nonlinear regime. We show that  it can be thought as measuring the memory of density and velocity fields to their initial conditions. This provides a clean definition of the validity of linear theory, which is shown to be much more restricted than usually recognized in the literature. We calculate the nonlinear propagator in RPT and compare to measurements in numerical simulations, showing remarkable agreement well into the nonlinear regime. We also show that N-body simulations require a rather large volume to recover the correct propagator, due to the missing large-scale modes. Our results for the nonlinear propagator provide an essential element to compute the nonlinear power spectrum in RPT.

\end{abstract}

\maketitle
s
\section{Introduction}

How long do perturbations ``remember" their initial state? What sets the validity of linear perturbation theory? The answer to these questions is important because, for example, measures of the power spectrum in the linear regime are being used~\cite{WMAP-SDSS} and planned to be used at higher redshifts~\cite{zeq3} to determine cosmological parameters. In the linear regime the fluctuations being observed contain the full information in the initial conditions~\footnote{Fundamentally, this is also true at the nonlinear level. The nonlinear evolution of collisionless dark matter is described by the Vlasov equation, which fully preserves information since the initial phase-space density {\em is conserved} along particle trajectories, the characteristic curves of the Vlasov-Poisson system. In practice we don't have full access to phase-space densities $f(\x,\p)$ and we try to learn about initial conditions from statistics of the evolved system. In this case whatever information is available is encoded in all the correlation functions of (moments with respect to $\p$ of) $f$. For a discussion of information in gravitational clustering from a different point of view see~\cite{infoP}.}. The breakdown of linear theory means that, unless we can somehow model deviations from it, only a very restricted range of scales is reliable for cosmological parameter estimation.

In this paper we study the propagator of density and velocity fields, the main ingredient that enters into a well-controlled extension of perturbation theory into the nonlinear regime (see our companion paper~\cite{paper1}, hereafter paper I). We shall see that the propagator measures the memory of initial conditions and its decay at high wavenumbers $k$ is nearly Gaussian with a characteristic scale that provides a clean definition of the scale at which linear theory breaks down. The end result of our study is an analytic expression for the fully nonlinear propagator of density and velocity fields that can be used in the formalism developed in paper I to calculate the nonlinear power spectrum~\cite{paper3},  and other statistics.

\begin{figure*}
\begin{center}
\begin{tabular}{cc}
{\includegraphics[width=0.5\textwidth]{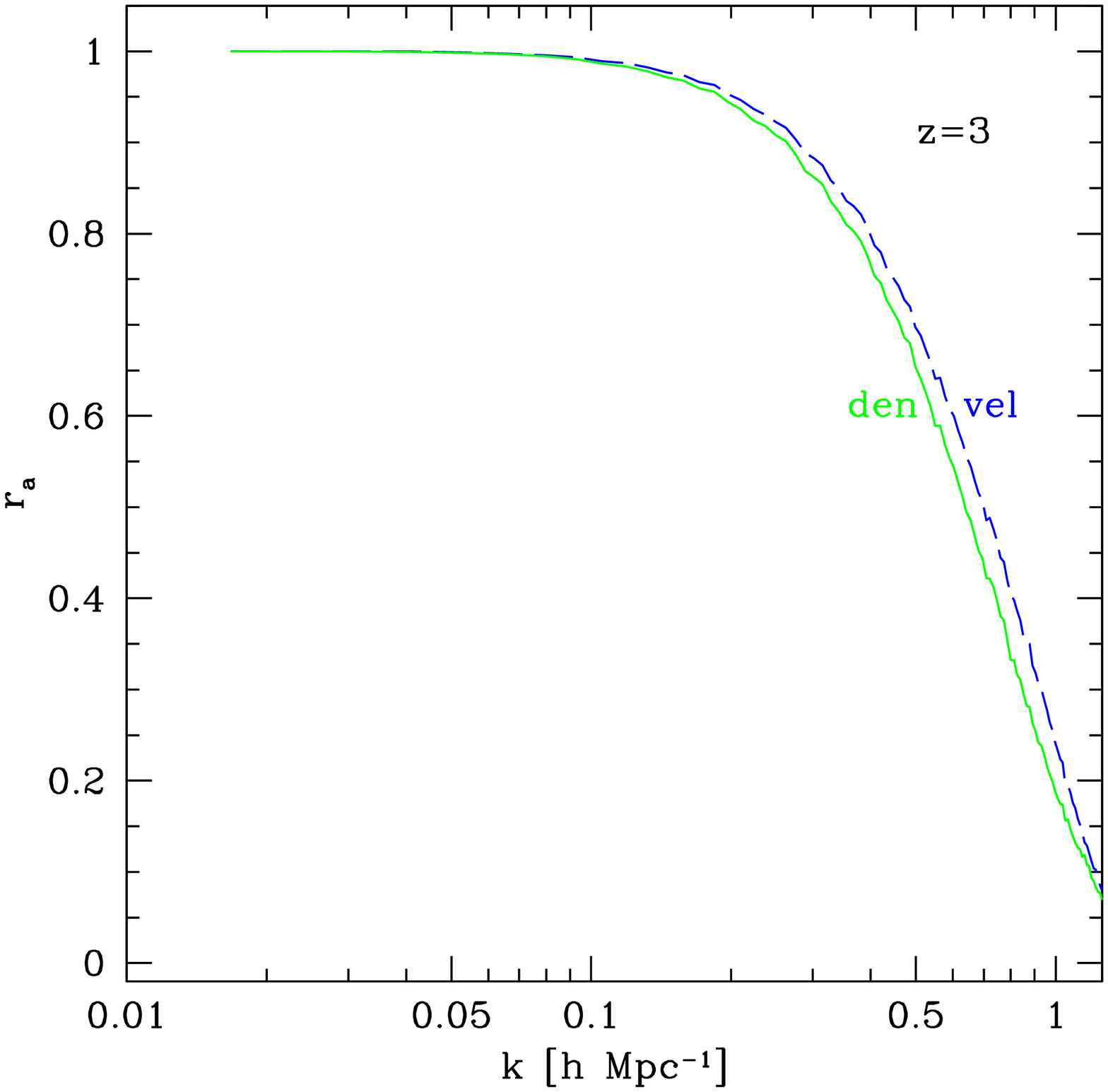}}&
{\includegraphics[width=0.5\textwidth]{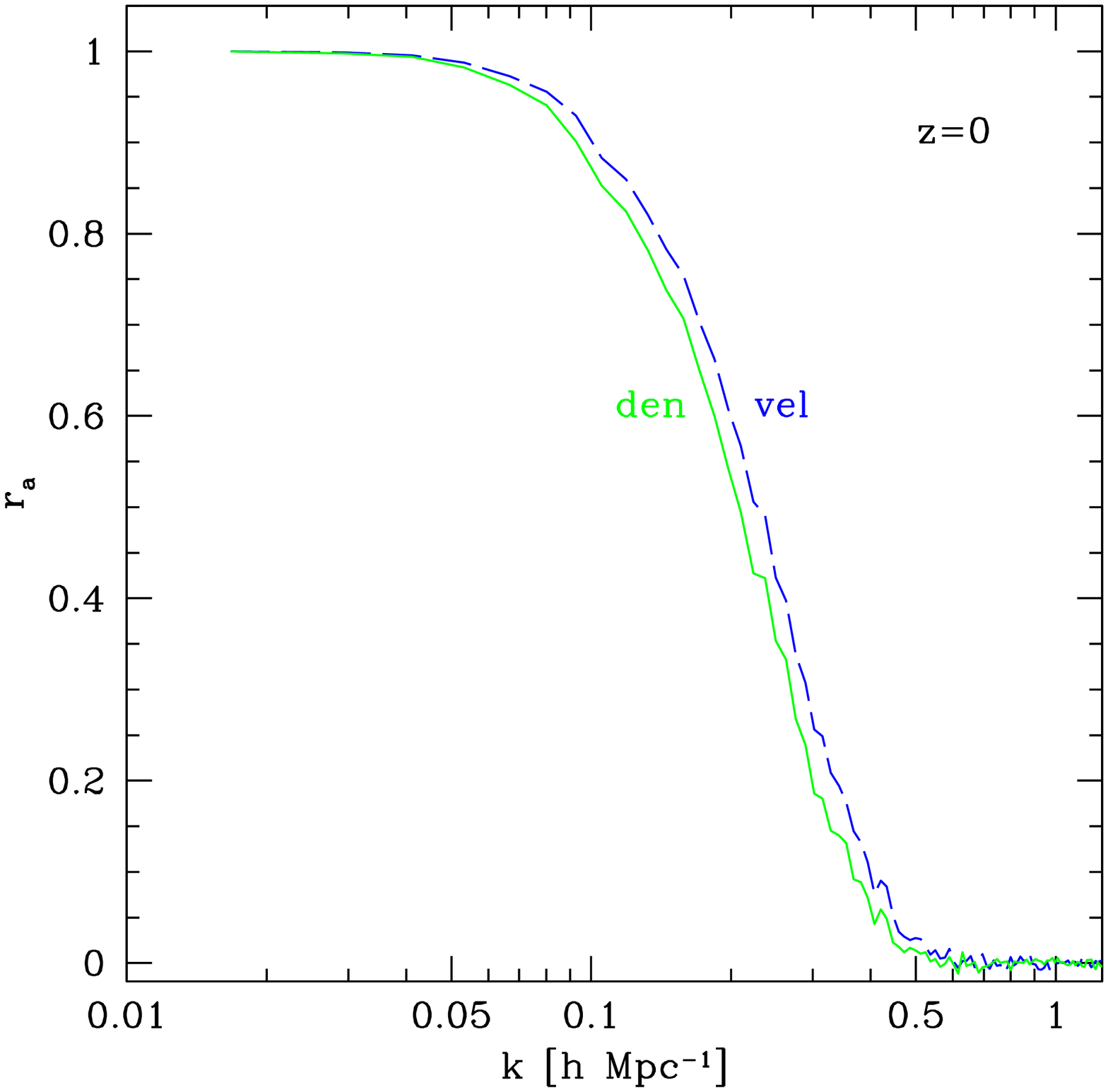}}
\end{tabular}
\caption{Cross-correlation coefficients between initial and final conditions $r_a$ as a function of wavenumber $k$ for density ($a=1$, solid lines) and velocity divergence ($a=2$, dashed lines) fields at redshift $z=3$ (left panel) and $z=0$ (right panel). Initial conditions are at redshift $z=35$.}
\label{figure1}
\end{center}
\end{figure*}

The most common criterion used in the literature to decide the validity of linear perturbation theory is to calculate the linear power spectrum and look for the scale at which the {\em rms} fluctuations are unity. This is intuitively reasonable, but in reality is very unreliable. The reason is that nonlinear corrections depend sensitively on the shape of the initial power spectrum (see e.g.~Fig.~12 in ~\cite{review}), not just on its amplitude. A better approach is to see how well does the final configuration of density and velocity fields looks like evolved from initial conditions by just pure linear theory. One statistical way of doing this is by calculating the cross-correlation coefficient $r_a$ between initial and final conditions as a function of scale,

\beq
r_a(k) \equiv \frac{X_a(k)}{\sqrt{P_a(k)P_0(k)}},
\label{crosscorr}
\eeq

where $\lexp \Psi_a(\k) \delta_0(\k') \rexp= X_a(k)\, \dD(\k+\k')$ is the cross-spectrum between initial (described by the density $\delta_0$, velocities are assumed in the linear growing mode) and final conditions (denoted by $ \Psi_a$, with  $a=1$ for density and $a=2$ for velocity divergence field, see Eq.(\ref{2vector}) for more details). In Eq.~(\ref{crosscorr}) $P_0$ denotes the initial power spectrum, and $P_a$ the final one. Figure~\ref{figure1} shows $r_a(k)$ at redshift $z=0,3$ for density and velocity fields. Note that, as expected, the cross-correlation decays slower for $z=3$ than $z=0$, indicating that linear theory is valid over a larger range of scales at $z=3$ than $z=0$. However, the characteristic scale of decay does not differ by as much as one would guess from the condition that the fluctuations be of order unity. Indeed, if one estimates the nonlinear scale $k_{\rm nl}$ from the linear spectrum, $4\pi k_{\rm nl}^3 P_L(k_{\rm nl})=1$, $k_{\rm nl}(z=0)=0.21 \kvecMpc$ and $k_{\rm nl}(z=3)=4.02 \kvecMpc$, a ratio of about 20, whereas the characteristic scale of decay of cross-correlations between initial and final fields (though it agrees with $k_{\rm nl}$ at $z=0$) only shifts by a factor of about $2-3$ at $z=3$. Therefore, the ``non-linear scale" can be misestimated by a large factor if $4\pi k_{\rm nl}^3 P_L(k_{\rm nl})=1$ is used at high redshifts. 

In practice, the cross-correlation coefficient between initial and final conditions is not entirely satisfactory as a measure of deviations from linear evolution, since the denominator in Eq.~(\ref{crosscorr}) is also sensitive to nonlinearities through $P_a(k)$, and this can lead to cancellations or enhacements. In particular, from Fig.~\ref{figure1} it appears that the density becomes nonlinear at slightly larger scales (lower $k$) than the velocities; however, this turns out to be incorrect, being more a reflection of the fact that nonlinear corrections to $P_a(k)$ are mostly positive for the density ($a=1$) and negative for the velocities ($a=2$). As we shall discuss in detail below, this problem can be avoided by the use of the propagator of density and velocity fields rather than the cross-correlation coefficient, the propagator $G_{ab}$ is defined by~\cite{Sco00,paper1}

\beq
G_{ab}(k,\eta)\ \delta_{\rm D}(\k-\k') \equiv \left\langle \frac{\delta \Psi_a(\k,\eta)}{\delta \phi_b(\k')}\right\rangle,
\label{NLProp}
\eeq
where $\Psi_a$ represents the final conditions and $\phi_a$ the initial conditions, and again $a=1$ corresponds to density and $a=2$ to velocity divergence fields. As we show below, for Gaussian initial conditions the propagator is proportional to the cross-correlation coefficient with a proportionality constant that gets rid of the dependence on $P_a(k)$, avoiding the problem with $r_a$. It is obvious from the definition in Eq.~(\ref{NLProp}) that if linear theory is valid the derivative yields a scale-independent result (the growth factor, for growing-mode initial conditions), and any dependence on scale is due to nonlinearities. As $k$ increases nonlinear effects become important and the final state does not ``remember" the initial one, driving the derivative to zero on average.  In this sense, the propagator can be though as a measure of the memory that perturbations have to their initial conditions. The characteristic scale of decay of the propagator is the cleanest measure of the scale where nonlinearities become important.

This paper is organized as follows. In the next section we review the basics of perturbation theory, see~\cite{review,paper1} for more details. Section~\ref{CalcProp} describes how we arrive at the analytic expression for the nonlinear propagator in RPT. Section~\ref{Measure} presents how to measure the propagator in numerical simulations and compares the results with RPT. Section~\ref{conclude} presents our conclusions.

\section{Equations of Motion and Perturbation Theory}

In this section we briefly review perturbation theory (PT) in a form useful to facilitate the process of resummation (see \cite{paper1} for more details). The equations of motion for the evolution of dark matter can be written in a compact way by introducing the two-component ``vector''

\beq
\Psi_a(\k,\eta) \equiv \Big( \delta(\k,\eta),\ -\theta(\k,\eta)/{\cal H} \Big),
\label{2vector}
\eeq
where the index $a=1,2$ selects the density or velocity components, with $\delta(\k)$ being the Fourier transform of the density contrast  $\delta (\x,\tau)=\rho(\x)/\bar \rho - 1$ and similarly for the peculiar velocity divergence $\theta\equiv\nabla\cdot{\bf v}$. ${\cal H}\equiv {d\ln a /{d\tau}}$ is the conformal expansion rate 
with  $a(\tau)$ the cosmological scale factor and $\tau$ the conformal time. The time variable $\eta$ is defined from the scale factor by 

\beq
\eta\equiv\ln a(\tau),
\label{timevariable}
\eeq
corresponding to the number of e-folds of expansion. Here we consider a cosmology where $\Omega_m=1$ and $\Omega_{\Lambda}=0$, see section~\ref{cosmodep} for the more general case. The equations of motion in Fourier space can then be written as (we henceforth use the convention that repeated Fourier arguments are integrated over)
\beq
\partial_{\eta} \Psi_a(\k,\eta) + \Omega_{ab} \Psi_b(\k,\eta) =
 \gamma_{abc}^{(\rm s)}(\k,\k_1,\k_2) \ \Psi_b(\k_1,\eta) \ \Psi_c(\k_2,\eta),
\label{eom}
\eeq
where 
\beq
\Omega_{ab} \equiv \Bigg[ 
\begin{array}{cc}
0 & -1 \\ -3/2 & 1/2 
\end{array}        \Bigg],
\eeq
and the symmetrized {\it vertex} matrix $\gamma_{abc}^{(\rm s)}$ describes the non linear interactions between different Fourier modes and is given by

\beqa
\gamma_{121}^{({\rm s})}(\k,\k_1,\k_2)&=&\delta_{\rm D}(\k-\k_1-\k_2) \  {(\k_1+\k_2) \cdot \k_1 \over{2 k_1^2}},
\label{ga121} \nonumber \\
\gamma_{222}^{({\rm s})}(\k,\k_1,\k_2)&=&\delta_{\rm D}(\k-\k_1-\k_2) \ {|\k_1+\k_2|^2 (\k_1 \cdot \k_2 )\over{2 k_1^2 k_2^2}},
\label{ga222}
\label{vertexdefinition}
\eeqa
$\gamma_{abc}^{(\rm s)}(\k,\k_i,\k_j)=\gamma_{acb}^{(\rm s)}(\k,\k_j,\k_i)$, and $\gamma$ is zero otherwise, $\delta_{\rm D}$ denotes the Dirac delta distribution. The formal integral solution to Eq. (\ref{eom}) is given by (see \cite{Sco98,Sco00,paper1} for a detailed derivation)

\beq
\Psi_a(\k,\eta) = g_{ab}(\eta) \ \phi_b(\k) + \int_0^\eta  d\eta' \ g_{ab}(\eta-\eta') \ \gamma_{bcd}^{(\rm s)}(\k,\k_1,\k_2)\ \Psi_c(\k_1,\eta') \Psi_d(\k_2,\eta'),
\label{eomi}
\eeq
where  $\phi_a(\k)\equiv\Psi_a(\k,\eta=0)$ denotes the initial conditions, set when the linear growth factor $a(\tau)=1$ and $\eta=0$. The {\em linear propagator} $g_{ab}(\eta)$ is given by 

\beq
g_{ab}(\eta) = \frac{{\rm e}^{\eta}}{5}
\Bigg[ \begin{array}{rr} 3 & 2 \\ 3 & 2 \end{array} \Bigg] -
\frac{{\rm e}^{-3\eta/2}}{5}
\Bigg[ \begin{array}{rr} -2 & 2 \\ 3 & -3 \end{array} \Bigg],
\label{prop}
\eeq 
for $\eta\geq 0$, whereas $g_{ab}(\eta) =0$ for $\eta<0$ due to causality, and $g_{ab}(\eta) \rightarrow \delta_{ab}$ as $\eta\rightarrow 0^{+}$. The {\em linear propagator} is the Green's function of the linearized version of Eq.~(\ref{eom}) and describes the standard linear evolution of the density and velocity fields from their initial state. 

An explicit expression for $\Psi_a(\k,\eta)$ can be written in the form of a series expansion

\beqa
\Psi_a(\k,\eta)=\sum_{n=1}^{\infty} \Psi^{(n)}_a(\k,\eta),
\label{seriesexp}
\eeqa
with 
\beq 
\Psi_a^{(n)}(\k,\eta)= \int \delta_{\rm D}(\k-\k_{1\ldots n})\ {\cal F}_{a a_1 a_2 \ldots a_n}^{(n)}(\k_1,\ldots,\k_n;\eta)\ \phi_{a_1}(\k_1) \ldots \phi_{a_n}(\k_n), 
\label{seriesol}
\eeq
where $\k_{1\ldots n} \equiv \k_1+ \ldots + \k_n$. Replacing Eq.~(\ref{seriesexp}) and~(\ref{seriesol}) into  Eq.~(\ref{eomi}) determines the recursion relations satisfied by the kernels ${\cal F}_{a a_1 a_2 \ldots a_n}^{(n)}$. 

In the standard form of Perturbation Theory (see~\cite{review} for a review), only the fastest growing mode is considered at each order. In contrast with this, Eq.~(\ref{seriesol}) gives the full time dependence, including all transients from initial conditions \cite{Sco98,paper1}. This is important since it plays a key role in allowing the process of resummation of the non linear propagator. See paper~I for a detailed discussion of this point.

It is worth noticing that so far we have not yet assumed any particular {\it kind} of initial conditions. From a physical point of view, the most interesting initial configurations are those where $\delta(\k,\eta=0)$ and $\theta(\k,\eta=0)$ are proportional Gaussian random fields, and thus we can write $\phi_a(\k)=u_a \delta_0(\k)$, with $\delta_0(\k)$ Gaussian distributed. Of particular importance when setting initial conditions in simulations are the {\em growing mode initial conditions}, for which $u=(1,1)$. Only the case $\phi_a(\k)=u_a \delta_0(\k)$ will be considered in this paper. Replacing this relation in Eq.~(\ref{seriesol}) we obtain a simplified version of it

\beq 
\Psi_a^{(n)}(\k,\eta)= \int \delta_{\rm D}(\k-\k_{1\ldots n})\, {\bar {\cal F}}_a^{(n)}(\k_1,\ldots,\k_n;\eta) \ \delta_0(\k_1) \ldots \delta_0(\k_n), 
\label{seriesolsimp}
\eeq
where $\bar {\cal F}_a^{(n)} \equiv {\cal F}_{a a_1 \ldots a_n}^{(n)} \,u_{a_1}\ldots u_{a_{2n}}$. The recursion relations for the kernels $\bar {\cal F}_a^{(n)}$ are given in \cite{paper1}.

\section{Calculating the Propagator}
\label{CalcProp}

\subsection{The Propagator as a Memory of Initial Conditions}
\label{memIC}

Non linear interactions lead to deviations from the linear evolution described by the first term in the r.h.s of Eq.~(\ref{eomi}). It is possible to interpret this effect as a modification of the linear propagator defined in Eq.~(\ref{prop}), this leads to a {\it propagator renormalization}. The study of this renormalized quantity, the non linear propagator, is the main goal of this paper. 

From its definition,  Eq.~(\ref{NLProp}), we see that the non linear propagator quantifies the dependence of the density and velocity fields on their initial values, in an {\it ensemble average} sense, and as we now show it has a very simple physical interpretation as a measure of memory of initial conditions. With the help of Eq.~(\ref{seriesexp}) we write Eq.~(\ref{NLProp}) as a series expansion

\beq
G_{ab}(k,\eta) = g_{ab}(\eta) + \sum_{n=2}^{\infty} \left\langle \frac{\delta \Psi^{(n)}_a(\k,\eta)}{\delta \phi_b(\k)}\right\rangle,
\label{PropagatorExpansion}
\eeq
where we have explicitly separated the linear part from the non-linear contributions. Now we want to make use of Eq.~(\ref{seriesol}) for $\Psi_a^{(n)}$, assuming Gaussian initial conditions. This implies that the statistical properties of the fields $\phi_a(\k)$ are completely characterized by the two-point correlator

\beq
\lexp \phi_a(\k) \ \phi_b(\k') \rexp = \delta_{\rm D}(\k+\k') \
u_a u_b\, P_0(k),
\label{2pt}
\eeq
where $P_0(k)$ denotes the initial power spectrum of density fluctuations. Therefore, for Gaussian initial conditions, only the odd terms in the series for $G_{ab}$ will give nonzero contributions after taking the functional derivative on $\Psi^{(n)}$ and performing the ensemble average. Replacing Eq.~(\ref{seriesol}) into Eq.~(\ref{NLProp}), we arrive at

\beq
G_{ab}(k,\eta) = g_{ab}(\eta) + \sum_{n=1}^{\infty} (2 n +1)!! \int \bar {\cal F}_{ab}^{(2n+1)}(\k,\k_1,-\k_1,\ldots,\k_n,-\k_n;\eta) P_0(k_1) \ldots  P_0(k_n),
\label{Gkernels}
\eeq
with
\beq
\bar {\cal F}_{ab}^{(2n+1)}(\k,\k_1,-\k_1,\ldots,\k_n,-\k_n;\eta) \equiv {\cal F}_{a b a_1 \ldots a_{2n}}^{(2n+1)}(\k,\k_1,-\k_1,\ldots,\k_n,-\k_n;\eta)\ u_{a_1}\ldots u_{a_{2n}} .
\label{Fbar}
\eeq

As pointed out in the introduction, the non linear propagator is closely related to the cross-correlation between final and initial conditions (assumed to be Gaussian). To see this, we multiply Eq.~(\ref{seriesol}) by $\phi_c(\k')$, and ensemble averaging assuming Gaussian initial conditions it follows that

\beq
G_{ab}(k,\eta)\,\langle\phi_b(\k)\phi_c(\k')\rangle=\langle\Psi_a(\k,\eta)\,\phi_c(\k')\rangle.
\label{Gcorr}
\eeq
Therefore, while in the non linear case it is not true anymore that $\Psi_a(\k,\eta)=G_{ab}(k,\eta)\phi_b(\k)$, since $G_{ab}$ is not the Green's function of Eq.~(\ref{eom}), the nonlinear propagator plays the role of a Green's function in two-point sense, $\langle \Psi_a \phi_b \rangle = G_{ac}\, \langle \phi_c\phi_b\rangle$. 
This is why $G_{ab}$ plays an essential role in calculating different correlation functions at the non linear level~\cite{paper1}. Also note that Eq.~(\ref{Gcorr}) makes explicit that $G_{ab}$ cannot depend upon the direction of $\k$, since both the initial and final configurations are statistically homogeneous and isotropic.

For the particular case when $\phi_b(\k)=u_b \delta_0(\k)$ and growing mode initial conditions, $u=(1,1)$, it follows that 

\beq
\delta_{\rm D}(\k+\k')\ [ G_{a1}(k,\eta)+G_{a2}(k,\eta) ]
=\frac{\langle\Psi_a(\k,\eta)\,\delta_0(\k') \rangle}{P_0(k)}.
\label{propcrosscorrelation}
\eeq
or, in terms of the cross-correlation coefficient, Eq.~(\ref{crosscorr})

\beq
G_{a1}(k,\eta)+G_{a2}(k,\eta) = {\rm r}_a(k,\tau) \sqrt{\frac{P_a(k,\tau)}{P_0(k)}}.
\label{relation}
\eeq

Therefore, the nonlinear propagator quantifies the memory of a density and velocity divergence field to its initial conditions as a function of scale and time,  a direct measure of the correlation between the final and initial configurations.

\begin{figure}[t!]
\begin{center}
\includegraphics[width=0.8\textwidth]{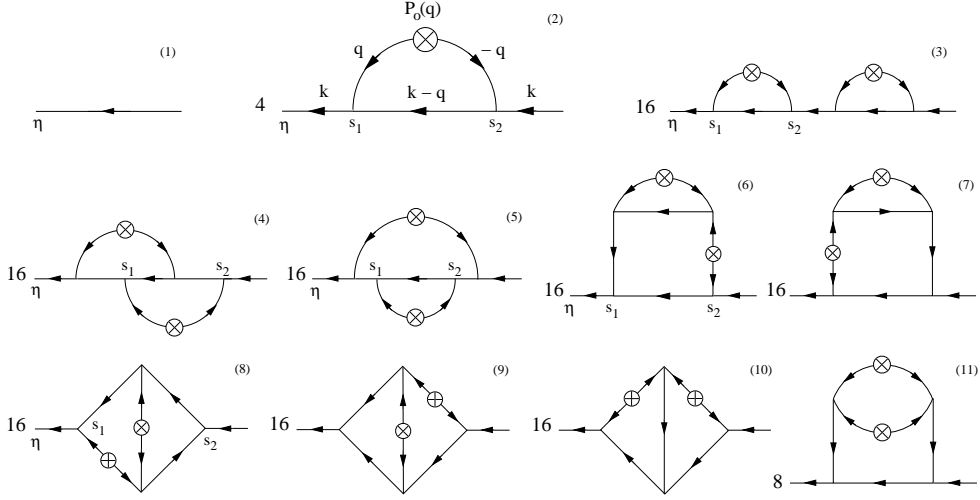}
\caption{Diagrams for the non linear propagator $G(k,\eta)$ up to two loops.}
\label{figure3}
\end{center}
\end{figure}

\subsection{Diagrammatic Representation}
\label{diagrammaticrepresentation}
In order to calculate analytically the terms in the series expansion of Eq.~(\ref{NLProp}), it is useful to resort to a diagrammatic representation (see paper I for more details). For Gaussian initial conditions, only the odd terms in the expansion for $G_{ab}$ are non zero, thus, we define

\beq
\delta G^{(n)}_{ab}(k,\eta)\ \delta_{\rm D}(\k-\k') \equiv \left\langle \frac{\delta \Psi^{(2n+1)}_a(\k,\eta)}{\delta \phi_b(\k')}\right\rangle,
\eeq
so that Eq.~(\ref{NLProp}) becomes, 
\beq
G_{ab}(k,\eta)=g_{ab}(\eta)+\sum_{n=1} \delta G^{(n)}_{ab}(k,\eta).
\label{volvioelaireacondicionado!!}
\eeq

In principle, an analytic expression for $\delta G_{ab}^{(n)}$ can be obtained, given the kernel $\bar{\mathcal F}^{(n)}_{ab}$, by integrating out Eq.~({\ref{Gkernels}). However, this path becomes very cumbersome with increasing $n$. In order to overcome this difficulty, and simplify the process of resummation, we developed in paper~I~\cite{paper1} a systematic approach in which the $n$th term in the series expansion of Eq.~(\ref{volvioelaireacondicionado!!}) is represented by a Feynman diagram with $n$ loops. Then, a set of rules enabled us to establish a one to one correspondence between diagrams and the corresponding integral contributing to $\delta G_{ab}^{(n)}(k,\eta)$. All the diagrams for the nonlinear propagator up to two loops are shown in Fig.~\ref{figure3}. Notice that the tree level diagram corresponds to the linear propagator $g_{ab}$ defined in Eq.(\ref{prop}).

A detailed description of the procedure to draw the diagrams, for both $\Psi^{(n)}_a$ and $\delta G^{(n)}_{ab}$, and the rules to write down the corresponding integrals is given in paper~I, here we only give a brief summary. The symbol $\otimes$ represents the initial power spectrum $u_a P_0(q) u_b$ [see Eq.~(\ref{2pt})], from where two lines emerge carrying opposite wavenumbers, say $\q$ and $-\q$. Each line finishes in a vertex, and represents a linear propagation $g_{ab}(s_i)$ from the initial conditions ($\eta=0$) to an arbitrary {\it time of interaction} $s_i$ at the vertex. At each vertex two modes interact, say $\k_1$ and $\k_2$, to give a nonlinear contribution to a third one, $\k$, with $\k=\k_1+\k_2$ (conservation of momentum). Each vertex in a diagram represents a vertex matrix $\gamma^{(s)}_{abc}(\k,\k_1,\k_2)$, defined in Eq.~(\ref{vertexdefinition}). The lines between vertices symbolize linear propagations between the corresponding times, e.g. $g(s_i-s_j)$. In addition, each diagram has an {\it outgoing} and an {\it incoming} line, identifiable with arrows. They represent, respectively, a linear propagator $g(\eta-s_1)$ and $g(s_{2n})$, where $s_1$ and $s_{2n}$ are the times corresponding to the first and last vertices of a general diagram with $n$ loops. Finally, all the intermediate wave vectors running through the loops are integrated over as well as all the time variables $s_j$, each between $[0,\eta]$.

\subsection{The One-Loop Propagator}
\label{onelooppropagator}

The first non-linear correction in Eq.~(\ref{volvioelaireacondicionado!!}) follows simply from applying these rules and is represented by diagram $2$ in Fig.~\ref{figure3}. It reads, 

\beqa
\delta G^{(1)}_{ab}(k,\eta)=4 \int_0^{\eta} ds_1 \int_0^{s_1} ds_2 \int d^3q \ \ g_{ac}(\eta-s_1)& \gamma^{(s)}_{cde}(\k,\q,\k-\q)& g_{df}(s_1)u_f\,g_{eg}(s_1-s_2) \nonumber \\ &
\gamma^{(s)}_{ghi}(\k-\q,-\q,\k)& g_{hj}(s_2)u_j\,g_{ib}(s_2)\,P_0(q) \ \ \ \ 
\label{OL1}
\eeqa
where the $\delta_{\rm D}$ in the definition of $\gamma^{(s)}$ in Eq.~(\ref{vertexdefinition}) has been integrated out, making explicit the conservation of momentum at each vertex. With the help of Eq.~(\ref{vertexdefinition}) and assuming growing mode initial conditions, is possible to write Eq.~(\ref{OL1}) as

\beq  
\delta G^{(1)}= \int_0^{\eta} ds_1 \int_0^{s_1} ds_2 \int_{-1}^1 dx \int_0^{\infty} 2\pi q^2 dq \ \ \frac{k x}{q}\Big(1-\frac{k x}{q}\Big)\,g(\eta-s_1)\,\Gamma\,g(s_1-s_2)\Gamma^{-1}\,g(s_2) e^{s_1+s_2} P_0(q)
\label{OL2}
\eeq
with
\beq
\Gamma \equiv \Bigg[ 
\begin{array}{cc}
x\, k/q+x\, q/k-2x^2 & 1-x\, q/k \\ 0 &  1-x\, k/q
\end{array}        \Bigg],
\eeq
and $x={\k}\cdot{\q}/k\, q$. Equation (\ref{OL2}) can be integrated out to give,

\beq
\delta G^{(1)}_{ij}(k,a)=\sum_m \, f^{(m)}_{ij}(k) \, a^m
\label{fns}
\eeq
where here $m$ takes the values  $m=\{3,2,1,1/2,-1/2,-3/2\}$ and $a(\tau)=\exp(\eta)$, Eq.~(\ref{timevariable}). It is possible to show that only $4$ functions $f^{(m)}_{ij}$ are independent, out of the $24$ in the set. We take these four to be, 
\beq
f\equiv\frac{5}{3}f^{(3)}_{11} \ \ ,\ \ g\equiv\frac{5}{2}f^{(-3/2)}_{11} \ \ ,\ \ h\equiv\frac{5}{2}f^{(1/2)}_{11} \ \ ,\ \ i\equiv\frac{5}{3}f^{(1)}_{11}.
\eeq
The normalization is such that in the large-$k$ limit $f,g,h,i \, \rightarrow -\frac{1}{2} k^2\sigma_v^2$, with $\sigma_v^2 \equiv \frac{1}{3} \int P_0(q) d^3q/q^2$, while they all vanish as $k \rightarrow 0$. Their explicit expressions are   

\beqa
f(k)&=&\int\frac{1}{504 k^3 q^5}
\left[6k^7q-79k^5q^3+50q^5k^3-21kq^7+\frac{3}{4}(k^2-q^2)^3(2k^2+7q^2)\ln \frac{|k-q|^2}{|k+q|^2}\right]P_0(q) \, d^3q, \nonumber \\
g(k)&=&\int\frac{1}{168k^3q^5}
\left[6k^7q-41k^5q^3+2k^3q^5-3kq^7+\frac{3}{4}(k^2-q^2)^3(2k^2+q^2)\ln\frac{|k-q|^2}{|k+q|^2}\right]P_0(q)\, d^3q,     \nonumber \\
h(k)&=&\int\frac{1}{24k^3q^5}
\left[6k^7q+k^5q^3+9kq^7+\frac{3}{4}(k^2-q^2)^2(2k^4+5k^2q^2+3q^4)\ln\frac{|k-q|^2}{|k+q|^2}\right]P_0(q) \,d^3q,   \nonumber    \\
i(k)&=&\int\frac{-1}{72k^3\,q^5}
\left[6k^7q+29k^5q^3-18k^3q^5+27kq^7+\frac{3}{4}(k^2-q^2)^2(2k^4+9k^2q^2+9q^4)\ln\frac{|k-q|^2}{|k+q|^2}\right] P_0(q) \, d^3q,  \nonumber \\    
\label{functionsofk}
\eeqa
Equation (\ref{fns}) can now be written in terms of these independent functions in a remarkably symmetric way

\beqa
\delta G_{11}^{(1)}(k,a)&=& \frac{3}{5} a \,\alpha(a)\ f(k) - \frac{3}{5} \beta(a)\ i(k) - \frac{2}{5} \gamma(a)\ h(k) + \frac{2}{5} a^{-3/2} \delta(a)\ g(k), \nonumber \\
\delta G_{12}^{(1)}(k,a)&=& \frac{2}{5} a \,\alpha(a)\ f(k) - \frac{2}{5} \beta(a)\ h(k) + \frac{2}{5} \gamma(a)\ h(k) - \frac{2}{5} a^{-3/2} \delta(a)\ f(k), \nonumber \\
\delta G_{21}^{(1)}(k,a)&=& \frac{3}{5} a \,\alpha(a)\ g(k) - \frac{3}{5} \beta(a)\ i(k) + \frac{3}{5} \gamma(a)\ i(k) - \frac{3}{5} a^{-3/2} \delta(a)\ g(k), \nonumber \\
\delta G_{22}^{(1)}(k,a)&=& \frac{2}{5} a \,\alpha(a)\ g(k) - \frac{2}{5} \beta(a)\ h(k) - \frac{3}{5} \gamma(a)\ i(k) + \frac{3}{5} a^{-3/2} \delta(a)\ f(k), 
\label{proponeloop1}
\eeqa
with 
\beqa
\alpha(a)=a^2-\frac{7}{5}a+\frac{2}{5}a^{-3/2},\,\beta(a)=\frac{3}{5}a^2-a+\frac{2}{5}a^{-1/2},\,\gamma(a)=\frac{2}{5}a^{2}-a^{1/2}+\frac{3}{5}a^{-1/2},\,\delta(a)=\frac{2}{5}a^{7/2}-\frac{7}{5}a+1.
\label{functionstime}
\nonumber
\eeqa

The low-$k$ limit of the one-loop propagator, $G^{(1)}=g+\delta G^{(1)}$, can be computed straightforwardly from Eqs.~(\ref{functionsofk}) and (\ref{proponeloop1}), reading~\cite{Sco00}

\beqa
G_{ab}^{(1)}(k,a) = g_{ab}(a) - (k \sigma_v)^2 a^3 \Bigg[
\begin{array}{cc} 61/350 & 61/525 \\ 27/50 & 9/25 \end{array}
\Bigg],
\label{lowklimit} 
\eeqa
where we kept only the fastest growing mode. Using Eqs.~(\ref{prop}) and (\ref{lowklimit}) we obtain for growing-mode initial conditions 

\beqa
G_\delta & \equiv & G_{11}+G_{12} \approx a\, \Big[1-\frac{61}{210}(k \sigma_v a)^2 \Big], 
\label{Gde} \\
G_\theta & \equiv & G_{21}+G_{22} \approx a\, \Big[1-\frac{9}{10}(k \sigma_v a)^2 \Big], 
\label{Gthe}
\eeqa
where we have defined the {\em density propagator} $G_\delta$ and the {\em velocity-divergence propagator} $G_\theta$ for growing mode initial conditions. We see that the one-loop corrections define a characteristic scale for which the propagator deviates significantly from its linear ($k$-independent) value. We shall come back to this important point below after we implement the process of resummation.

\subsection{Resummation in the large-$k$ limit}
\label{resummationlargek}

\begin{figure}[t!]
\begin{center}
\includegraphics[width=0.9\textwidth]{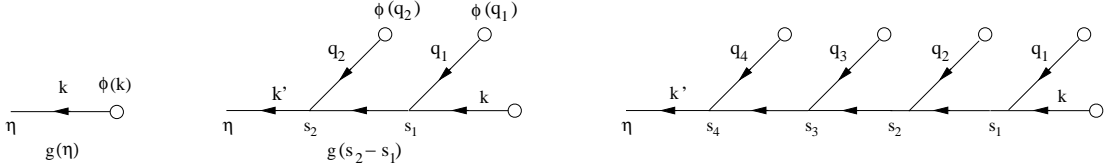}
\caption{Tree diagrams for $\Psi_a(\k,\eta)$, up to 4 vertices, that lead to the non linear propagator's resummable subset.}
\label{tree}
\end{center}
\end{figure}

As we discussed in paper~I, the non linear propagator plays an essential role in renormalized perturbation theory, since it allows a well-defined expansion for the power spectrum and higher-order correlation functions into the nonlinear regime. Therefore, one of the main objectives of this paper is to find an adequate analytical description for $G_{ab}$. In order to achieve this, we would  need to sum the infinite number of terms defining $G_{ab}$ in Eq.~(\ref{volvioelaireacondicionado!!}), equivalent to adding up an infinite number of  diagrams, shown in Fig.~\ref{figure3} up to two loops.

Although it is very likely that this resummation cannot be done exactly,  we shall see that reasonable physical arguments can be put forward to simplify the task enough so that it becomes doable. The first step is to realize that since the propagator measures how memory to initial conditions is lost, the next best thing to having an exact calculation of it, is to include in the calculation those processes that lead to the {\em slowest} decay, or memory loss, since they are the ones that establish the dominant behavior. 

Different ``processes" first appear at two-loops, where there is more than one diagram contributing. 
The two-loop diagrams in Fig.~\ref{figure3} can be organized according to how many interactions (vertices) take place along the principal path of the diagram (that connects the beginning to the end without going through initial conditions). For example diagrams 3,4,5 have all four interactions along the principal path, diagrams 8,9,10 have three and diagrams 6,7,11 have only two interactions along the principal path. Diagrams that have all interactions along the principal path are special in the sense that the principal path interacts only with waves coming directly from the initial conditions in their growing mode. In addition, in the high-$k$ limit each of these interactions results in a wave still in the growing mode, since as we show below in this case $\gamma_{abc}u_b u_c \propto u_a$ and this keeps repeating itself interaction after interaction along the principal path. For the other diagrams, on the contrary, the interactions outside the principal path do not preserve the linear growing mode in the high-k limit (since they involve momenta that cannot be compared to $k$). Therefore,  {\em when resummed} these should be suppressed compared to the resummation of the ones where all interactions happen along the principal path, since they should cross-correlate less with initial conditions. One exception to this is when the time between final and initial conditions is small enough, in which case there is no time for decorrelation and the distinction we made is not significant. 

The dominant subset as defined above can be resummed exactly in the high-$k$ limit. In general, the subset has $(2n-1)!!$ diagrams with $n$ loops, and it gives rise to diagram 1 (tree level), diagram 2 (one loop) and diagrams 3, 4 and 5 (two loops) in Fig.~\ref{figure3}. The infinite number of diagrams in this subset originate in tree diagrams for $\Psi_a(\k,\eta)$, as described in paper~I, with a very simple topology. They all start in two Fourier modes that evolve linearly from the initial conditions in growing mode, with $u_a=(1,1)$, interact (represented by a vertex), and give rise to a third mode. This mode, in turn, evolves and interacts with a fourth mode that had grown linearly until that time, i.e. with no previous interaction. This process continues until the final time $\eta$, with $2n$ interactions needed to describe the $n$-loop diagrams. In Fig.~\ref{tree} we show the trees with 0, 2 and 4 vertices that will originate the {\em tree}, {\em one-loop} and {\em two-loop} diagrams respectively. {\em In contrast with the rest of the diagrams, the ones belonging to this subset have the simplest ramification possible and thus most direct connection to the initial conditions}. As discussed above, we expect this subset to be the dominant, although we cannot prove it rigorously since we could not sum up the remaining subset and show it is indeed subdominant.

The expression for $\Psi_a^{(n)}$, relevant to compute $\delta G_{ab}^{(n)}$ in Eq.~(\ref{volvioelaireacondicionado!!}), can be written recursively as,

\beq
\Psi^{(n)}_a(\k',s)=2\int_0^s ds_n\,g_{ab}(s-s_n) \gamma^{(s)}_{bcd}(\k',\q_n,\k+\q_{n-1\ldots1})\Psi_c^{(1)}(\q_n,s_n)\Psi_d^{(n-1)}(\k+\q_{n-1\ldots1},s_n)
\label{rec1}
\eeq
with $\k'=\k+\q_{n\ldots1}$ and

\beq
\Psi^{(1)}_a(\q,s)=g_{ab}(s)\,\phi_b(\q)=e^{s}\,u_a\,\delta_0(\q)
\label{linearmode}
\eeq
The factor of $2$ in Eq.~(\ref{rec1}) is because all the branchings in Fig.~\ref{tree} are asymmetric if $n>2$. This should be replaced by 1 for $n=2$. Next, we need to take the functional derivative of Eq.~(\ref{rec1}) with respect to $\phi_b(\k)$. An important condition, in order to recover the desired diagrams, is that the functional derivative must be taken only on any of the two linear modes $\Psi^{(1)}$ that precede the very first vertex in Fig.~\ref{tree}. In other words, the derivative must be iterated through Eq.~(\ref{rec1}) until $n=2$. After using  Eq.~(\ref{rec1}) for $\Psi_a^{(2n+1)}$, taking the derivative as described and using Eq.~(\ref{linearmode}), we arrive at

\beqa
\frac{\delta \Psi^{(2n+1)}_a(\k',\eta)}{\delta \phi_b(\k)}=2^{2n}\int_0^{\eta} ds_{2n} \int_0^{s_{2n}} ds_{2n-1} \ldots \int_0^{s_2} ds_1 \left[g(\eta-s_{2n})\,\bar\gamma^{(s)}(\k+\q_{2n\ldots1},\q_{2n},\k+\q_{2n-1\ldots1})\right. \nonumber \\ 
\left. g(s_{2n}-s_{2n-1})\ldots g(s_2-s_1)\,\bar\gamma^{(s)}(\k+\q_1,\q_1,\k)\,g(s_1)\right]_{ab} \delta_0(\q_{2n})\ldots\delta_0(\q_1) \, e^{s_{2n}+\ldots+s_1},
\label{rec2}
\eeqa
where $\bar\gamma$ is a $2\times2$ matrix defined as $\bar\gamma^{(s)}_{ac}\equiv\gamma^{(s)}_{abc}\, u_b$, and $\k'=\k+\q_{2n\ldots1}$. In the large-k limit, where the external momentum $\k$ is much larger than any internal $\q_i$, we can expand $\bar\gamma$ to leading order, with the help of Eq.~(\ref{vertexdefinition}), as

\beq
\bar\gamma^{(s)}_{ab}(\k+\q_{n\ldots1},\q_n,\k+\q_{n-1\ldots1})\approx\frac{1}{2} \frac{k}{q_n} x_n \, \delta_{ab}
\label{VE}
\eeq
where $x_n= \k \cdot \q_n / k\,q_n$, and $\delta_{ab}$ is the $2\times2$ identity matrix. Equation~(\ref{VE}) shows that in the high-$k$ limit $\gamma_{abc}u_b u_c \propto u_a$ at each vertex for this subset of diagrams, as discussed above. A great simplification of Eq.~(\ref{rec2}) follows from the fact that $\bar\gamma$ is proportional to the identity matrix in the large-k limit. Replacing Eq.~(\ref{VE}) into Eq.~(\ref{rec2}), and using that $g(u-s)g(s-z)=g(u-z)$, we obtain

\beqa
\left\langle\frac{\delta \Psi^{(2n+1)}_a(\k',\eta)}{\delta \phi_b(\k)} \right\rangle =g_{ab}(\eta) \frac{{(a(\eta)-1)}^{2n}}{(2n)!}\int d^3q_{2n}\ldots d^3q_1\,\frac{k^{2n}}{q_{2n}\ldots q_1}\,x_{2n}\ldots x_1\,\langle\delta_0(\q_{2n})\ldots\delta_0(\q_1)\rangle
\label{rec3}
\eeqa
where we have integrated the $2n$ time variables according to,

\beqa
\int_0^{\eta} ds_{2n} \int_0^{s_{2n}} ds_{2n-1} \ldots \int_0^{s_2} ds_1 \, e^{s_{2n}+\ldots+s_1} = \frac{{(a(\eta)-1)}^{2n}}{(2n)!}
\nonumber
\eeqa
with $\eta=\ln(a)$. The ensemble average of the $2n$ Gaussian fields leads to $(2n-1)!!$ equal contributions to Eq.~(\ref{rec3}), corresponding to all different pairings of the fields $\delta_0$. In addition, for any given pairing there are $n$ equal contributions, each given by

\beq
\int\int \frac{k^2}{q_i q_j}\, x_i x_j \, \langle \delta_0(\q_i)\delta_0(\q_j)\rangle \, d^3q_i d^3q_j = \frac{-k^2}{3} \int \frac{P_0(q)}{q^2} d^3 q \equiv -k^2 \sigma_v^2
\label{ec4}
\eeq
Replacing Eq.~(\ref{ec4}) to the $n$th power into Eq.~(\ref{rec3}), and the result into Eq.(\ref{volvioelaireacondicionado!!}) we get

\beq
G_{ab}(k,a)=g_{ab}(a)+ g_{ab}(a) \sum_{n=1} \frac{(2n-1)!!}{(2n)!} \left[ -k^2 \sigma_v^2 (a-1)^2 \right]^n
\eeq
This series can be summed up to give
\beq
G_{ab}(k,a)=g_{ab}(a) \exp(-k^2 \sigma_v^2 (a-1)^2/2),
\label{largeKresult}
\eeq
which shows that we expect the propagator to decay as a Gaussian in the large-$k$ limit.

\subsection{The Propagator in Renormalized Perturbation Theory}
\label{propRPT}

We would like now to extend the result in Eq.~(\ref{largeKresult}) to obtain an analytic result for the nonlinear propagator than can be used at any $k$. The Gaussian dependence on $k$ is also recovered, as shown  in paper~I, for the {\em exact} density propagator $G_{\delta}=G_{11}+G_{12}$  in the Zel'dovich approximation, which has a Gaussian dependence on $k$, {\em for all} $k$, $G^{{\rm ZA}} _{\delta}(k,a)= a\, {\rm e}^{-\frac{1}{2} k^2 \sigma_v^2 a^2}$ for growing-mode initial conditions. 

Based on these considerations we assume that the nonlinear propagator has a nearly Gaussian behavior, with a characteristic scale of decay that can be then determined from knowing the behavior of the propagator for low-$k$, the one-loop result in Eq.~(\ref{proponeloop1}). 

In order to do this and obtain a well defined nonlinear propagator we first rewrite $\delta G^{(1)}$ in Eq.~(\ref{proponeloop1}) in a way similar to the linear propagator. That is, we group the contributions to each component of $\delta G^{(1)}$ in two terms, one ``growing'' (with an overall factor $a$), and one ``decaying'' (with an overall $a^{-3/2}$). The only guidelines for defining each term are that the most growing contribution (i.e. $\alpha$) must belong to the growing part. In addition, both terms must be always negative, in particular in the limits of large $a$ (for all scales) and large-$k$ (for all times). This guarantees the nonlinear propagator will decay in the long-time or large-$k$ limit.

From their definition in Eq.~(\ref{functionsofk}), it is easy to check that $f,g,i$ are always negative and diverge monotonously to $-\infty$ as $k$ increases, as $-k^2 \sigma^2_v(\tau)/2$. Moreover, $h(k)>f(k)>g(k), i(k)$ for all $k$. Provided with this and the definitions of $\alpha$, $\beta$, $\gamma$ and $\delta$ in Eq.~(\ref{functionstime}), it is possible to conclude that the only feasible way to satisfy our requirements is to group things in the following way,

\beqa
g_{11}(a)+\delta G_{11}^{(1)}(k,a)&=& \frac{3}{5} \, a + \frac{2}{5} \, a^{-3/2} + \frac{3}{5} \, a \, [\,\alpha(a) \, f(k) - \beta_g(a) \, i(k)\,] + \frac{2}{5} \, a^{-3/2} \, [\,\delta(a) \, g(k) - \gamma_d(a) \, h(k)\,] \nonumber \\
g_{12}(a)+\delta G_{12}^{(1)}(k,a)&=& \frac{2}{5} \, a - \frac{2}{5} \, a^{-3/2} + \frac{2}{5} \, a \, [\,\alpha(a) \, f(k) - \beta_g(a) \, h(k)\,] - \frac{2}{5} \, a^{-3/2} \, [\,\delta(a) \, f(k) - \gamma_d(a) \, h(k)\,] \nonumber \\
g_{21}(a)+\delta G_{21}^{(1)}(k,a)&=& \frac{3}{5} \, a - \frac{3}{5} \, a^{-3/2} + \frac{3}{5} \, a \, [\,\alpha(a) \, g(k) + \gamma_g(a) \, h(k)\,] - \frac{3}{5} \, a^{-3/2} \, [\,\delta(a) \, g(k) + \beta_d(a) \, i(k)\,]\nonumber \\
g_{22}(a)+\delta G_{22}^{(1)}(k,a)&=& \frac{2}{5} \, a + \frac{3}{5} \, a^{-3/2} + \frac{2}{5} \, a \, [\,\alpha(a) \, g(k) - \frac{3}{2} \gamma_g(a) \, i(k)\,] + \frac{3}{5} \, a^{-3/2} \, [\,\delta(a) \, f(k) - \frac{2}{3} \beta_d(a) \, h(k)\,] \nonumber,  \\
\label{proponeloop}
\eeqa
 where we have redefined $\beta$ and $\gamma$ in Eq.~(\ref{functionstime}) through  $a\,\beta_g(a)=a^{-3/2} \beta_d(a)=\beta(a)$, and $a\,\gamma_g(a)=a^{-3/2} \gamma_d(a)=\gamma(a)$. Next we regard  Eq.~(\ref{proponeloop}) as a nearly Gaussian expansion of the full result and write

\begin{eqnarray}
g_{11}+\delta G^{(1)}_{11}\,\rightarrow\,G_{11}(k,a)&=&\frac{3}{5} \, a \, {\rm e}^{(\alpha(a) f(k) - \beta_g(a) i(k))} + \frac{2}{5} \, a^{-3/2} \, {\rm e}^{(\delta(a) g(k) - \gamma_d(a) h(k))} \nonumber ,\\
g_{12}+\delta G^{(1)}_{12}\,\rightarrow\,G_{12}(k,a)&=&\frac{2}{5} \, a \, {\rm e}^{(\alpha(a) f(k) - \beta_g(a) h(k))} - \frac{2}{5} \, a^{-3/2} \, {\rm e}^{(\delta(a) f(k) - \gamma_d(a) h(k))}  \nonumber ,\\
g_{21}+\delta G^{(1)}_{21}\,\rightarrow\,G_{21}(k,a)&=&\frac{3}{5} \, a \, {\rm e}^{(\alpha(a) g(k) + \gamma_g(a) h(k))} - \frac{3}{5} \, a^{-3/2} \, {\rm e}^{(\delta(a) g(k) + \beta_d(a) i(k))} \nonumber ,\\
g_{22}+\delta G^{(1)}_{22}\,\rightarrow\,G_{22}(k,a)&=&\frac{2}{5} \, a \, {\rm e}^{(\alpha(a) g(k) - (3/2) \gamma_g(a) i(k))} + \frac{3}{5} \, a^{-3/2} \, {\rm e}^{(\delta(a) f(k)-(2/3) \beta_d(a) h(k))}.
\label{model3one}
\end{eqnarray}
This is our prediction for the nonlinear propagator in RPT that we will contrast against N-body simulations below. Note that there are no free parameters, given an initial spectrum and cosmological parameters we can fully predict $G_{ab}(k,\eta)$. Before we do so, however, we must take into account how our calculation is changed when we go beyond the $\Omega_m=1$, $\Omega_\Lambda=0$ model we have considered so far.

\subsection{Dependence on $\Omega_m$ and $\Omega_\Lambda$}
\label{cosmodep}

The dependence on cosmological parameters ignored so far can be taken into account, to a very good approximation. First, if we redefine

\beq
\Psi_a(\k,\eta) \equiv \Big( \delta(\k,\eta),\ -\theta(\k,\eta)/{\cal H} f \Big),
\label{2vector2}
\eeq
where $f \equiv  {\rm d}\ln D_+/{\rm d}\ln a$, with $D_+$ the linear growth factor in the appropriate cosmology, and
\beq
\eta\equiv\ln D_+(\tau),
\label{timevariable2}
\eeq
the equations of motion take the same form as in Eq.~(\ref{eom}) with the only change~\cite{SCFFHM98}

\beq
\Omega_{ab} \equiv \Bigg[ 
\begin{array}{cc}
0 & -1 \\ -3\Omega_m/2f^2 & 1/2 
\end{array}        \Bigg],
\label{Omega2}
\eeq

These results are so far exact. The problem in Eq.~(\ref{Omega2}) is that both $\Omega_m$ and $f$ are functions of time, therefore the Laplace transform solution in Eq.~(\ref{eomi}) is not correct anymore. Although the growing mode will be correct, due to the change in Eqs.~(\ref{2vector2}-\ref{timevariable2}), the decaying mode is not, since in general it is given by the Hubble constant $H$ which does not scale as $D_+^{-3/2}$ in the general case. However it is well known in the standard formulation of perturbation theory, that the dependence of the perturbative solutions in the values of $\Omega_m$ and $\Omega_{\Lambda}$ is extremely weak, once the linear growth factors have been scaled out~\cite{Bouchet,Bernardeu,review}. In other words, almost all of the information about the cosmological parameters is encoded in the linear growth factor $D_{+}(\tau)$. The reason is that during most of the time evolution $\Omega_m/f^2 \approx 1$~\cite{SCFFHM98}. Therefore, when we compare the propagator in RPT against simulations for the $\Lambda$CDM model in the next section, we will simply replace $a(\tau)$ in Eq.~(\ref{model3one}) by the corresponding linear growth factor $D_+(\tau,\Omega_m,\Omega_{\Lambda})$.

\section{Measuring the Propagator in Numerical Simulations}
\label{Measure}

We now describe how to measure the propagator in numerical simulations. The simulations we use here correspond to three different box sizes, $L_{\rm box}=100$, $239.5$, $479 \Mpc$ with a number of particles $N_{\rm par}=200^3$, $256^3$, $512^3$, respectively. They all correspond to the $\Lambda$CDM model with $\Omega_m=0.3$, $\Omega_\Lambda=0.7$ and $\sigma_8=0.9$ at redshift $z=0$. The small and medium simulations have an input power spectrum with shape parameter $\Gamma=0.21$, the large simulation has instead an input power spectrum that includes the full transfer function with $\Omega_b=0.04$ in baryons. We have 24 realizations of the small simulation (see~\cite{karim} for a description of these), and only one realization of the medium and large simulations. In the following we present results from the large simulation, except for section~\ref{voldep} where we explore the dependence on the simulation volume.

The propagator has never been measured in a numerical simulation before. Here we explore two methods, one based on an implementation of the numerical derivative involved in its definition, Eq.~(\ref{NLProp}), the other based on the relationship between the propagator and the cross-correlation coefficient, Eq.~(\ref{relation}). We show that both methods give the same result.

\subsection{Computing the Functional derivative numerically}
\label{fdn}

According to Eq.~(\ref{NLProp}), the propagator involves a functional derivative of the final conditions with respect to the initial conditions, which formally is given by

\beq
\frac{\delta \Psi_a(\k)}{\delta \phi_b(\k')} = \lim_{\epsilon \rightarrow 0}\  \frac{\Psi_a[\phi_b(\k)+\epsilon \, \dD(\k-\k')] - \Psi_a[\phi_b(\k)]}{\epsilon}.
\eeq

In order to measure $G_{ab}$ for a given Fourier mode $\k_i$, based on this definition we would need to evolve two simulations whose initial configurations of density and velocity fields are equal, except at the value of the Fourier coefficient $\phi_b(\k_i)$, where they differ by a small amount $\epsilon$ (but consistent with Gaussian initial conditions). Next, the difference of the final Fourier coefficients $\Psi_a(\k_i)$ must be divided by the initial difference $\epsilon$. This needs to be done several times for each Fourier mode, to obtain an expectation value, and repeated for each scale of interest. Computationally, this clearly represents a nearly impossible task.

Instead, we proceed based on an assumption that resembles the notion of ``ergodicity'', namely, that all Fourier modes $\k_i$, within a bin of wave vectors of magnitude around  $|\k|$, can be thought as different realizations of the same Fourier mode. This is the same idea one uses in calculating the power spectrum from a single realization, here also making use of the fact that $G_{ab}$ depends only on the magnitude of $\k$ due to statistical homogeneity and isotropy (see section~\ref{memIC}), as does the power spectrum. Therefore, given one realization of the simulation we compute the difference between $\Psi_a(\k_i)$ and the value of the field $\Psi_a$ at another Fourier mode $\k_j$, within the same bin determined by $|\k|$. Then, we divide the result by their difference at the initial conditions, $\phi_b(\k_i)-\phi_b(\k_j)$. Finally, we average over all pairs of modes within each bin. 

All the simulations we use were set up using growing mode initial conditions, where $\delta=\theta=\delta_0$ initially. As a result of this, it is only possible to measure a ``directional functional derivative'' $\langle \delta \Psi_a / \delta \phi_b \rangle \cdot u_b=G_{a1}+G_{a2} $, which following Eqs.~(\ref{Gde}-\ref{Gthe}) above we shall call the density ($G_\delta=G_{11}+G_{12}$) and velocity divergence ($G_\theta=G_{21}+G_{22}$) propagators.

In summary, given the Fourier coefficients of the final density ($a=1$) or velocity divergence fields ($a=2$), and the initial conditions $\delta_0(\k)$, we estimate the non linear propagators as

\beqa
G_{a1}(k,\eta)+G_{a2}(k,\eta)=\frac{1}{N(N-1)} \sum_{{\bf k}_i} \sum_{{\bf k}_j \neq {\bf k}_i } \frac{ \Psi_a({{\bf k}_i,\eta})- \Psi_a({{\bf k}_j},\eta)}{\delta_0({{\bf k}_i})- \delta_0({{\bf k}_j})},
\label{funcderivative}
\eeqa
where the sum runs over pairs of Fourier modes in the  bin of magnitude $|\k|$. The result is real because the sum involves pairs of Fourier modes with opposite directions.

\subsection{Computing the Propagator from the cross-correlation}
\label{crosscorrelationmemoryfunction}

\begin{figure*}
\begin{center}
\begin{tabular}{cc}
{\includegraphics[width=0.5\textwidth]{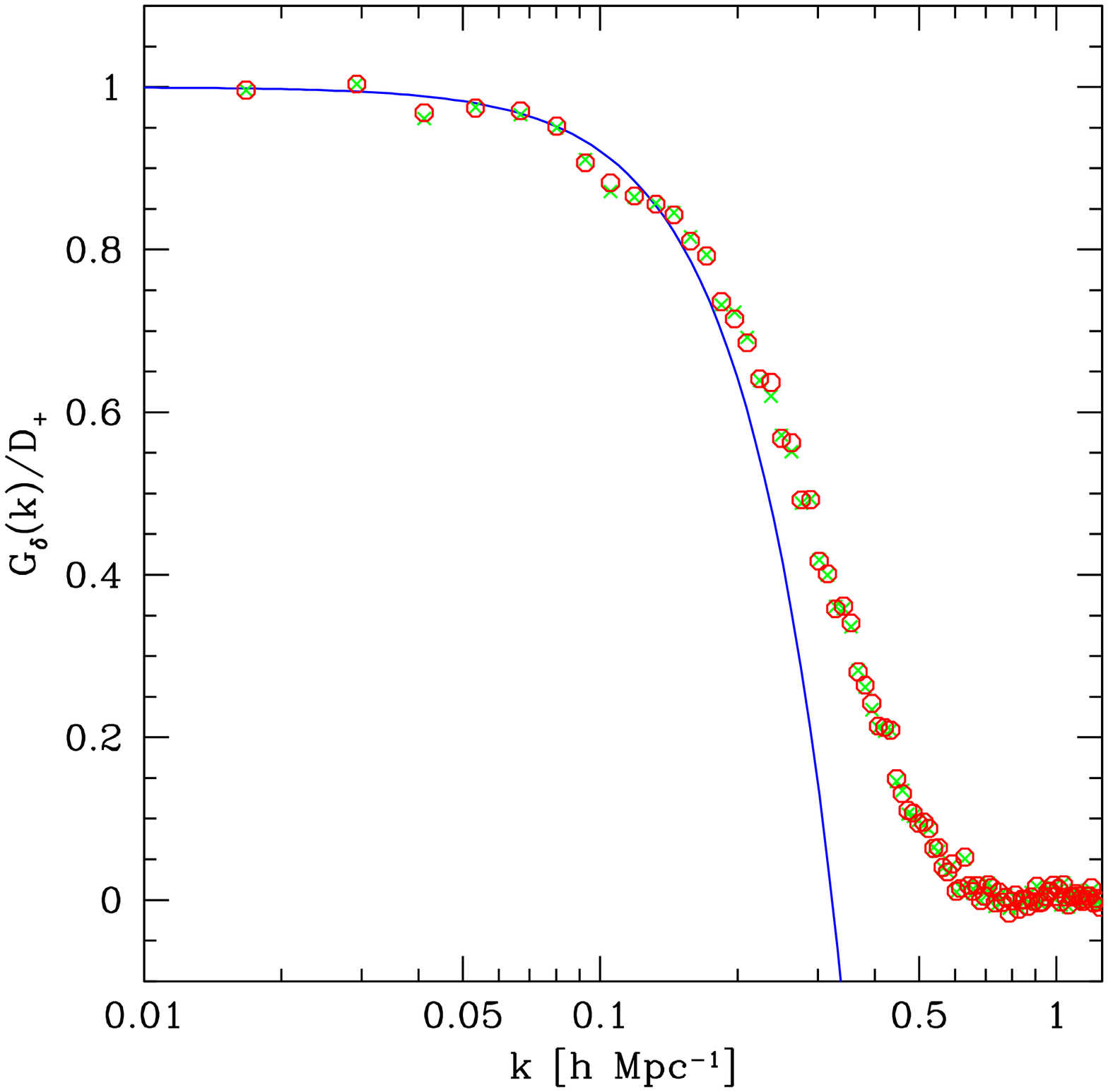}}&
{\includegraphics[width=0.5\textwidth]{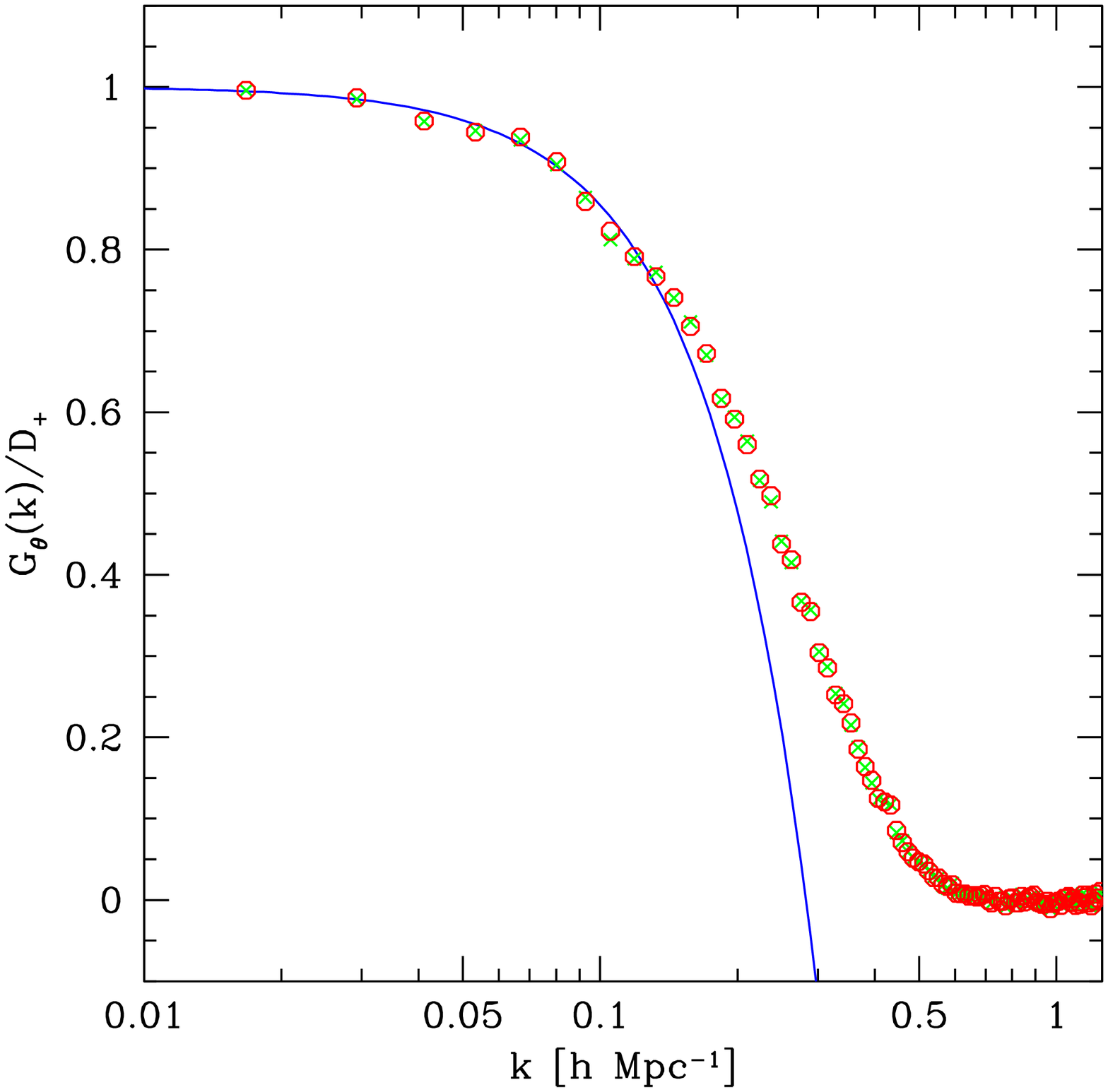}}
\end{tabular}
\caption{The density (left panel) and velocity divergence (right panel) propagators at redshift $z=0$ from initial conditions at $z=5$ (corresponding to $D_+=4.68$). The symbols represent measurement in numerical simulations, from implementation of the functional derivative (crosses) and from the relation to the cross-correlation coefficient (circles). The solid lines show the predictions of one-loop PT, Eq.~(\ref{proponeloop1}).}
\label{DerVsX}
\end{center}
\end{figure*}

An alternative approach is to take advantage of the relationship in Eq.~(\ref{propcrosscorrelation}) between the propagator and the cross-correlation coefficient. This leads to a straightforward method to measure the density and velocity propagators directly from simulations as

\beqa
G_{a1}(k,\eta)+G_{a2}(k,\eta)=\frac{1}{P_0(k)}\frac{1}{N} \sum_{{\bf k}_i} \Psi_a({{\bf k}_i,\eta})\delta_0({-{\bf k}_i}),
\label{crosscorrmethod}
\eeqa
where the sum runs over Fourier modes in the $|\k|$ bin. Notice that the two methods proposed to measure $G_{a1}+G_{a2}$ in Eqs~(\ref{funcderivative}) and (\ref{crosscorrmethod}) are intrinsically very different. One is rooted in the definition of the non linear propagator as a functional derivative while the other exploits the relation between $G_{ab}$ and the cross-correlation coefficient. Nevertheless, Fig.~\ref{DerVsX} shows that the two methods give essentially the same answer. Since the cross-correlation method is much faster (being of order $N$ as opposed to $N^2$) we shall use that in the following. The only limitation of Eq.~(\ref{crosscorrmethod}) is that it assumes Gaussian initial conditions, however, as long as the initial redshift is large enough this should not be a problem, as Fig.~\ref{DerVsX} shows for $z_{\rm initial}=5$.

Figure~\ref{DerVsX} also includes a comparison with the one-loop propagators from Eq.~(\ref{proponeloop1}), showing that in the low-$k$ limit the one-loop result does very well in describing the decay of the nonlinear propagators. We also see that the one-loop propagator becomes negative (and very quickly  large as $k$ increases), which is unphysical given its physical interpretation as measuring the memory to initial conditions. This makes clear the main point of RPT discussed in paper~I: unless the propagator is resummed to recover the right behavior as $k \rightarrow \infty$, the loop expansion in standard PT gives rise to large negative contributions which hinder the convergence of PT as nonlinear scales are probed. However, from Fig.~\ref{DerVsX} we see that the one-loop corrections are not too far off from describing the full decay of the propagator, therefore an approximate resummation as described  in section~\ref{propRPT}, stands a good chance of providing a good description.

\subsection{The Propagator in RPT vs. N-Body Simulations}

\begin{figure*}
\begin{center}
\begin{tabular}{cc}
{\includegraphics[width=0.5\textwidth]{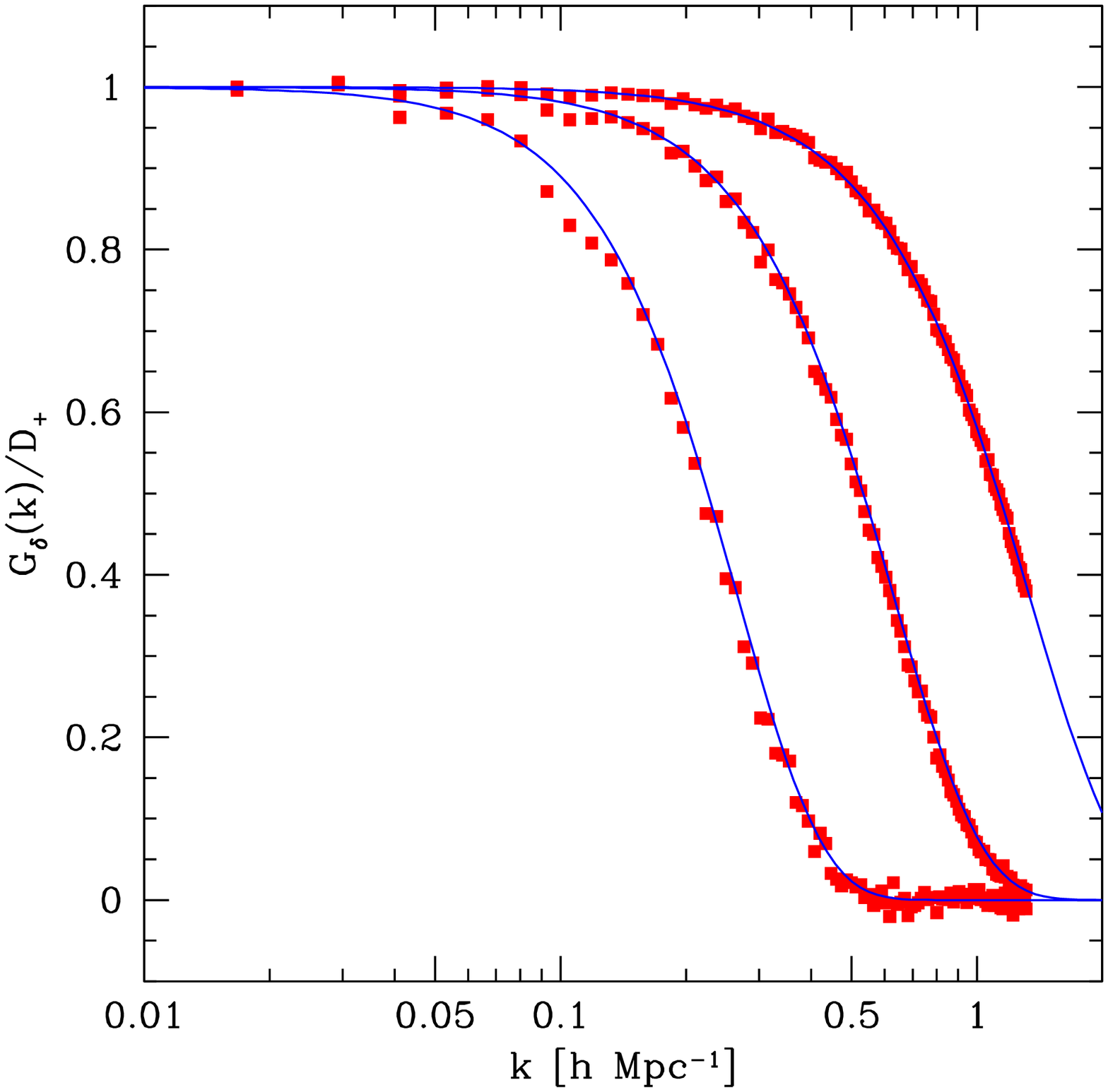}}&
{\includegraphics[width=0.5\textwidth]{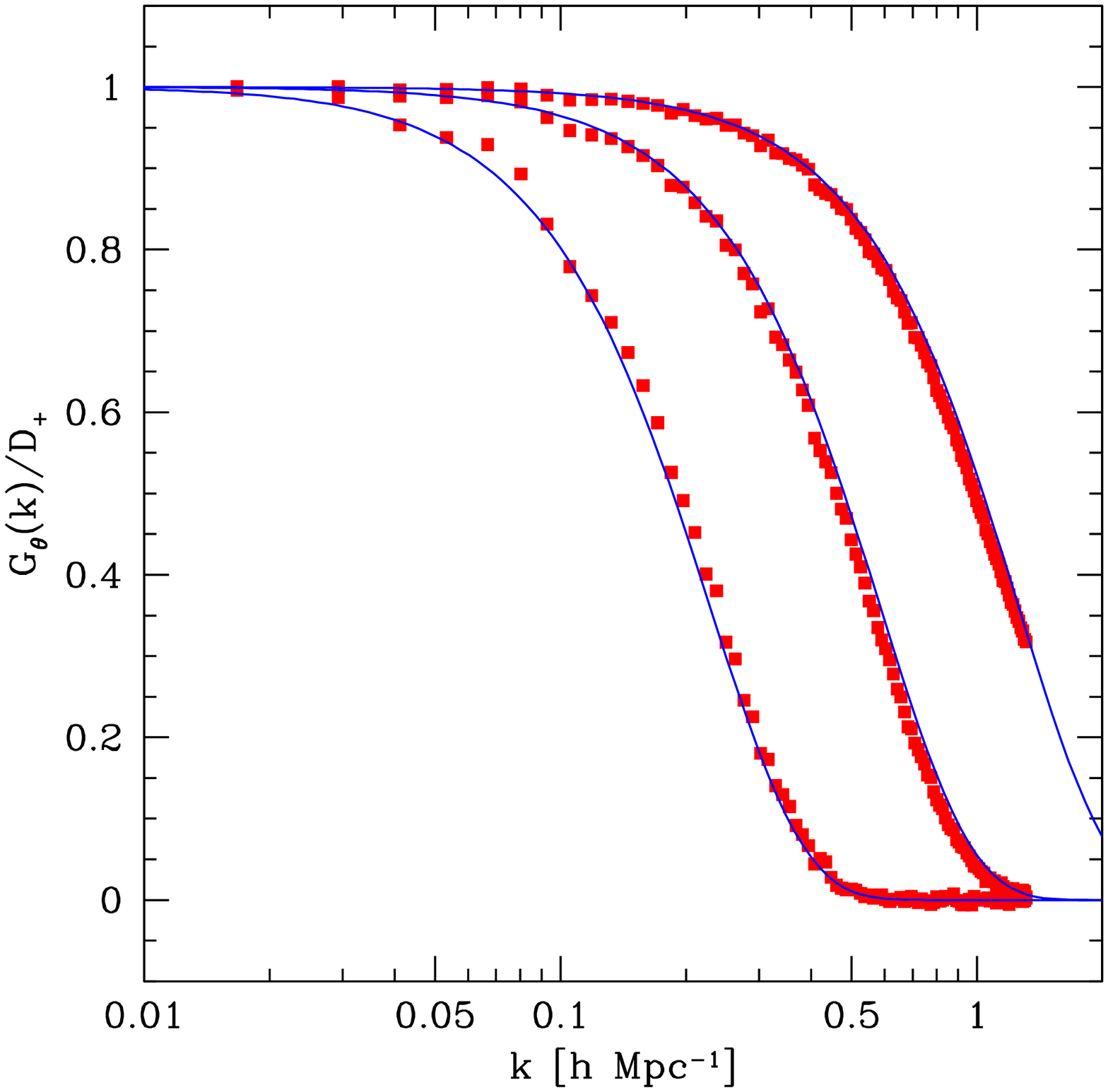}}
\end{tabular}
\caption{Predictions for the propagator of density (left panel) and velocity divergence (right panel) from Eq.~(\ref{model3one}) against measurements in N-body simulations. The three cases correspond (from left to right in each panel) to $z=0,2,5$, with the initial conditions at $z_{\rm initial}=35$. The predictions shown in solid lines {\em have no free parameters}.}
\label{RPTvsNB}
\end{center}
\end{figure*}

Figure \ref{RPTvsNB} shows the comparison between the nonlinear propagator from Eq.~(\ref{model3one}) and measurements in the large numerical simulation, for various redshifts $z=0,2,5$ (corresponding to linear growth factors $D_+=28,11.8,5.9$) from initial conditions at $z_{\rm initial}=35$. The left panel shows the density propagator, whereas the right panel corresponds to the velocity divergence propagator. From this we see that, despite the approximations involved in the resummation process, Eq.~(\ref{model3one}) describes the dependence of the non linear propagator with scale and time remarkably well {\em without introducing any free parameters}. The reason why we can predict the behavior of the nonlinear propagator down to scales in the nonlinear regime can be summarized as,

\begin{itemize}

\item[i)] The one-loop correction already describes the onset of the decay of the propagator very well (Fig.~\ref{DerVsX}), 

\item[ii)] In the high-$k$ and long-time limit one expects the subset of diagrams with all interactions along the principal path to dominate over the rest. These can be resummed exactly, Eq.~(\ref{largeKresult}), leading to Gaussian behavior,

\item[iii)] In order to match these two results and thus complete the resummation for all $k$, we make the assumption that the one-loop result is a nearly Gaussian expansion of the renormalized propagator. Then, as discussed in section~\ref{propRPT} there is a unique way to recover the latter from it's one-loop expansion satisfying the right asymptotics, leading to Eq.~(\ref{model3one}). 

\end{itemize}

\subsection{Dependence of Propagators on the Simulation Volume}
\label{voldep}

\begin{figure*}
\begin{center}
\begin{tabular}{cc}
{\includegraphics[width=0.5\textwidth]{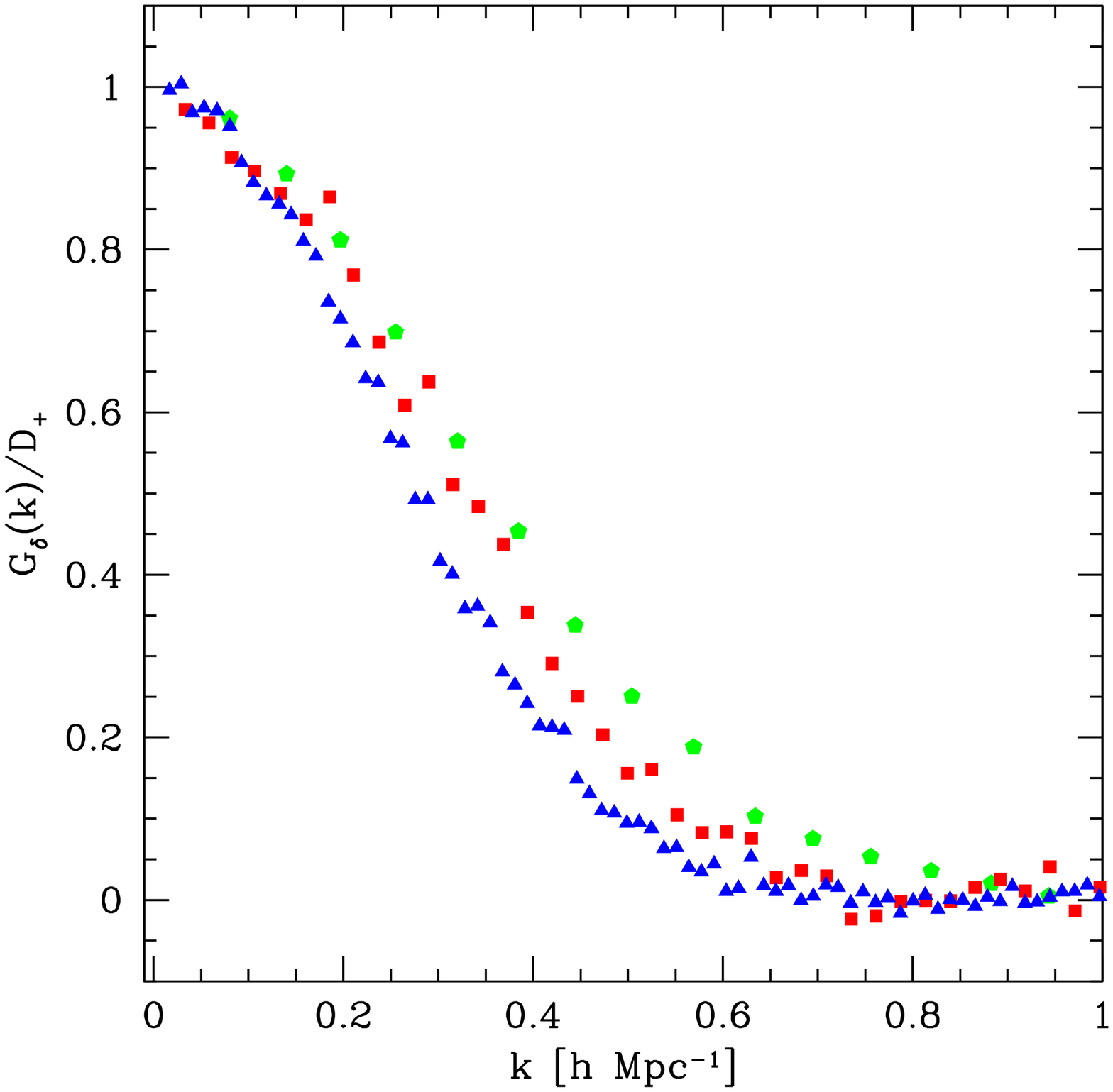}}&
{\includegraphics[width=0.5\textwidth]{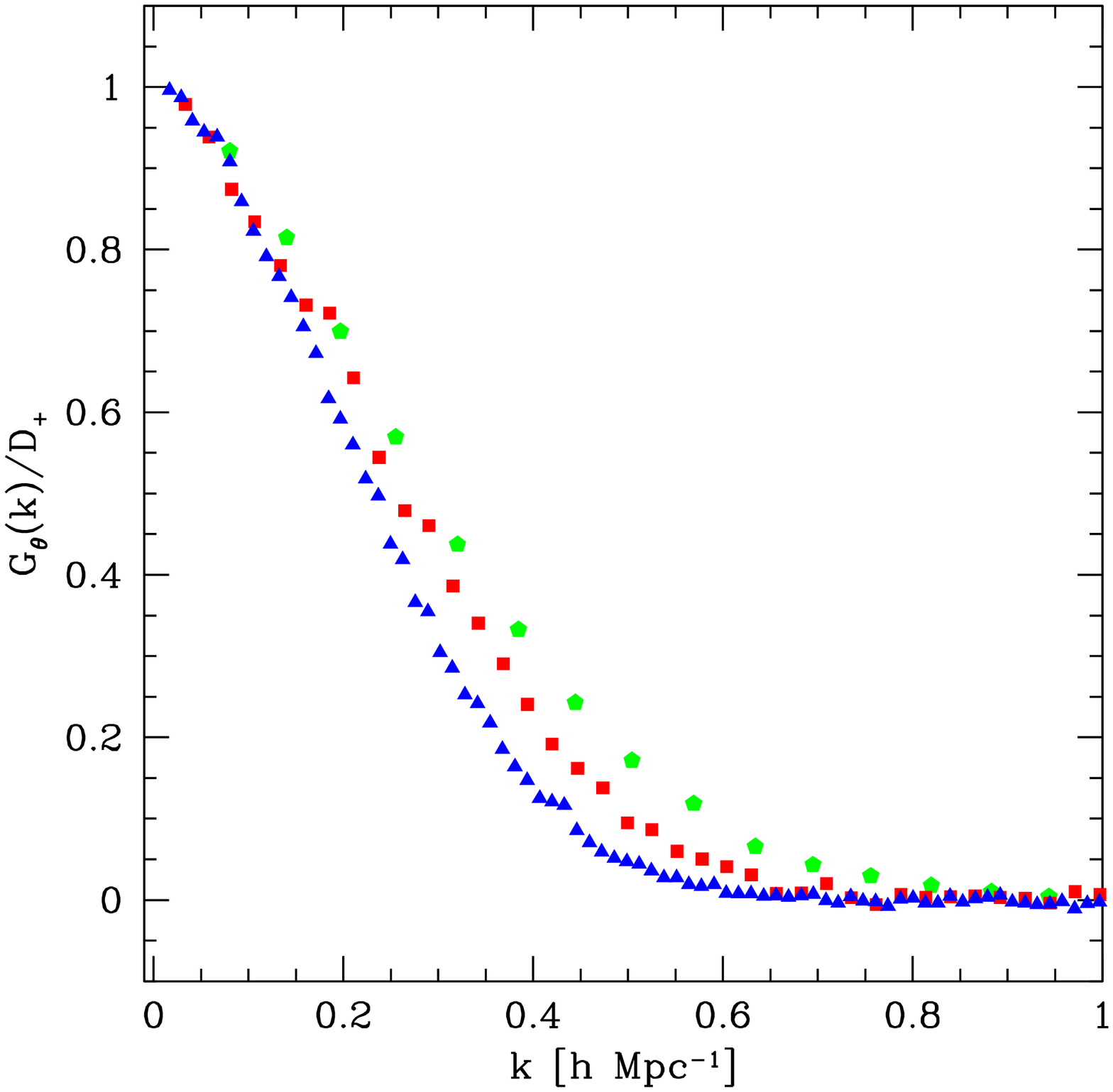}}
\end{tabular}
\caption{The dependence of the measured density (left) and velocity divergence (right) propagators on the volume of the simulation used. The symbols denote measurements of the propagator at $z=0$ from initial conditions at $z_{\rm initial}=5$ for different simulation volumes, $L_{\rm box}=100 \Mpc$ (pentagons), $L_{\rm box}=239.5 \Mpc$ (squares), $L_{\rm box}=479 \Mpc$ (triangles).}
\label{PropLbox}
\end{center}
\end{figure*}

There is one important aspect of the prediction from the RPT propagator that we have ignored so far, namely, its  dependence on the large-scale modes. This is important because numerical simulations artificially truncate large-scale modes due to their finite volume. The predictions from Eqs.~(\ref{proponeloop1}),~(\ref{largeKresult}), and~(\ref{model3one}) indicate that the dependence on the large-scale modes must be significant, since the characteristic scale of decay is determined by $\sigma_v^{-1}$, which is quite sensitive to the smallest wavenumber $k_{\rm box}$ available in a simulation volume. In fact, the analytic predictions presented already in all the figures use the exact value $k_{\rm box}=2\pi/L_{\rm box} =0.013 \kvecMpc$ corresponding to the large simulation box. From Eqs.~(\ref{Gde}-\ref{Gthe}) we see that one can define the characteristic scale of decay from the low-$k$ expansion of the density and velocity propagators as

\beqa
\label{kdecayden}
k_\delta(z)&=&\sqrt{105/61}\ \sigma_v^{-1}(z), \\
\label{kdecayvel}
k_\theta(z)&=&\sqrt{5/9}\ \sigma_v^{-1}(z), \\
\label{svz}
\sigma_v^2(z) &=& (4\pi/3) D_+^2(z) \int_{k_{\rm box}}^\infty P_0(k) dk,
\eeqa
which is roughly the scale at which $G_\delta/D_+$ and $G_\theta/D_+$ have respectively dropped to $e^{-1/2}\approx 0.6$. 

Figure~\ref{PropLbox} shows the dependence of the propagators on the simulation volume, confirming that there is a significant dependence, in agreement with the estimates from Eqs.~(\ref{kdecayden}-\ref{kdecayvel}). A larger volume leads to a stronger decay due to the fact that more modes are present and the increased interactions lead to faster loss of memory of the initial conditions. A similar situation holds for the cross-correlation coefficient between initial and final conditions presented in Fig.~\ref{figure1}. In other words, the larger the volume allows interactions among more modes which leads to a larger scatter of the final densities and velocities about the linearly extrapolated initial conditions, which suppresses $r_a$.

Why is such a strong dependence on the simulation volume not seen in the power spectrum measurements? The reason is that in a finite volume the propagator decays slower, leading to a fictitiously wider validity of linear perturbation theory, but the truncation of low-$k$ modes also suppresses the mode-coupling terms that should enhance the power, therefore these two effects compensate each other. {\em However}, that means that the dependence of the power spectrum on cosmological parameters will be incorrect at the transition scale, since these two contributions have a different dependence on them (their ratio being proportional to the linear power spectrum) and therefore cannot completely cancel. 

The importance of this finite-volume effect is stronger for steep spectra, in fact one should correct the $L_{\rm box}=479 \kvecMpc$ measurements in Fig.~\ref{PropLbox} by a shift of about $4\%$ to the right according to Eqs.~(\ref{kdecayden}-\ref{svz}) to take into account the difference between the initial spectrum of this simulation as opposed to the $L_{\rm box}=100,239.5 \Mpc$ ones, but this does not alter the conclusions significantly.

Moreover, using Eqs.~(\ref{kdecayden}-\ref{svz}) one concludes that the decay length of the propagators in the $L_{\rm box}=479 \kvecMpc$ case is still off by about $4\%$ compared to the infinite volume case that will show a stronger decay. An important lesson from this is that simulations designed to study the high-redshift Universe that generally use smaller boxes can misestimate nonlinear effects due to lack of large-scale modes very easily, as the requirement to get the decay scale of the propagator correctly  {\em is independent of redshift} and, for example, to get $k_\delta,k_\theta$ accurate to $10\%$ requires a simulation of about $L_{\rm box}=240 \Mpc$, which can be challenging for high-redshift numerical studies.

\begin{figure}[t!]
\includegraphics[width=0.5\textwidth]{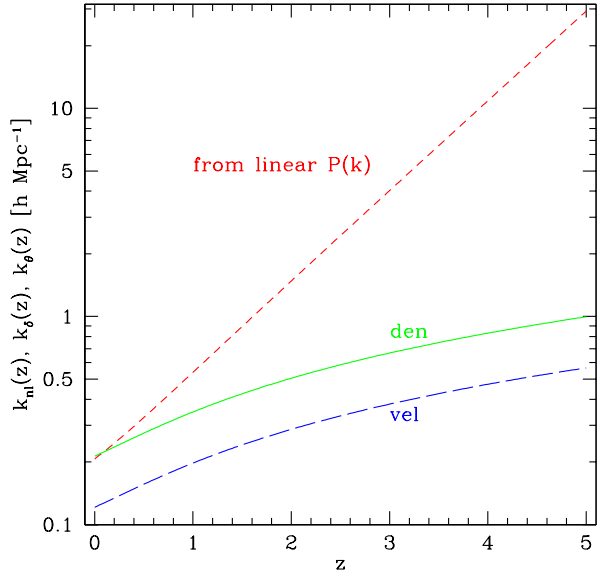}
\caption{Comparison between the nonlinear scale defined from the linear power spectrum $k_{\rm nl}(z)$ (short-dashed lines) and those derived from the characteristic scale of decay of the nonlinear propagator $k_\delta(z)$ (solid lines), $k_\theta(z)$ (long-dashed lines) given by Eqs.~(\ref{kdecayden}-\ref{kdecayvel}). At high redshift $k_{\rm nl}(z)$ significantly underestimates the importance of nonlinearities.}
\label{NLscale}
\end{figure}

\section{Conclusions}
\label{conclude}

In paper~I~\cite{paper1} we developed renormalized perturbation theory (RPT), in which nonlinear evolution in the growth of large-scale structure is  represented by Feynman diagrams constructed in terms of three objects: the initial conditions (e.g. perturbation spectrum), the vertex (describing non-linearities) and the propagator (describing linear evolution). We showed that loop corrections to the linear power spectrum organize themselves into two classes of diagrams: one corresponding to mode-coupling effects, the other to a renormalization of the propagator. The nonlinear propagator that results  quantifies the deviation from linear evolution of individual Fourier modes, and allows a well-defined perturbation theory that can probe the nonlinear regime. 

In this paper we studied the propagator in detail and showed that its decay into the nonlinear regime can be thought as measuring the ``memory'' of perturbations to their initial conditions, being directly proportional to the cross-correlation coefficient between final and initial configurations.  We developed an analytical description of the nonlinear propagator by summing its perturbation series using physically motivated approximations. The results are in remarkable agreement with measurements in numerical simulations all the way into the nonlinear regime.  We also described how to measure the propagator in numerical simulations using two very different algorithms that gave consistent answers. The analytic results show that the propagator is rather sensitive to the finite volume of simulations, and this is indeed in agreement with measurements in simulations of different volume. The requirement of getting the right dependence of the propagator on scale turns out to be a stringent one for N-body simulations, particularly at high redshift where typically smaller boxes are used. 

Coming back to the questions raised in the introduction about the validity of linear perturbation theory, we can give a quantitative answer for a robust definition of the nonlinear scale by looking at the characteristic scale of decay of the propagator of density and velocity divergence fields $k_\delta,k_\theta$ given by Eqs.~(\ref{kdecayden}-\ref{kdecayvel}). Figure~\ref{NLscale} shows the evolution of $k_\delta,k_\theta$ as a function of redshift and compared to the nonlinear scale $k_{\rm nl}$ defined from the linear power spectrum, $4\pi k_{\rm nl}^3 P_L(k_{\rm nl})=1$. Although the different definitions agree for the density field at $z=0$, at high redshift the nonlinear scale $k_{\rm nl}$ typically used in the literature can be more than an order of magnitude off from $k_\delta,k_\theta$, which we argued here is the cleanest definition of at what scale nonlinearities become important.

Since the non linear propagator plays an essential role in the calculation of  correlations functions, the next obvious step is to use the results presented here to calculate the non linear power spectrum. In the framework of RPT, once the nonlinear propagator is known, the power spectrum is given as a mode-coupling series~\cite{paper1}. Although in principle many terms are needed in order to cover a large range of scales, the study of how baryon wiggles are affected by nonlinear evolution requires only a few terms~\cite{paper3}. Already, the dependence of the propagator on scale found here provides a simple estimate of the scale at which  baryon wiggles should be significantly suppressed (remaining only as $e^{-1/2} \approx 0.6$ of their linear amplitude over a smooth component), given by $k_\delta(z),k_\theta(z)$ in Fig.~\ref{NLscale}.

\acknowledgments

We thank G. Gabadadze, A. Gruzinov, D.W. Hogg, S. Pueblas, and E. Sefusatti,  for useful discussions. The small-box numerical simulations used here where run at the NYU Beowulf cluster supported by NSF grant PHY-0116590. We thank K. Benabed for his help on that. The medium and large simulations were carried out by the Virgo Supercomputing Consortium using computers based at the Computing Centre of the Max-Planck Society in Garching and at the Edinburgh parallel Computing Centre. The data are publicly available at {\tt http://www.mpa-garching.mpg.de/NumCos}.

\end{document}